# The Effects of Substrate Temperature on Indium Gallium Nitride Nanocolumn Crystal Growth

*S. Keating, M.G.Urquhart, D.V.P.McLaughlin and J.M.Pearce\**

*Department of Mechanical and Materials Engineering, Queen's University*

\* Corresponding author: 60 Union Street, Kingston, Ontario, K7L 3N6 Canada, ph: 613-533-3369, email: pearce@me.queensu.ca

**Abstract**

Indium gallium nitride films with nanocolumnar microstructure were deposited with varying indium content and substrate temperatures using plasma-enhanced evaporation on amorphous $SiO_2$ substrates. FESEM and XRD results are presented, showing that more crystalline nanocolumnar microstructures can be engineered at lower indium compositions. Nanocolumn diameter and packing factor (void fraction) was found to be highly dependent on substrate temperature, with thinner and more closely packed nanocolumns observed at lower substrate temperatures.

**Introduction**

The growing environmental concerns on the combustion of fossil fuels are driving a renewed interest in developing third-generation photovoltaic (PV) devices, which use less raw materials and have higher efficiencies than conventional silicon PV (1,2). One of the most promising techniques for improving PV efficiency is the use of multi-junction cells to absorb a wider energy range of incident photons (3). Indium gallium nitride ($In_xGa_{1-x}N$) is an ideal material candidate for multi-junction cells as its range of band gaps covers the solar spectrum: 0.65eV to 3.4eV depending on the relative indium content, *x* (4-7). Recent investigations into InGaN single-junction photovoltaics have reported successful results in achieving significant photoresponses from experimental devices (8,9). In addition to band gap engineering, PV device performance can be improved by engineering the microstructure of the material to increase the optical path length and provide light trapping. For this purpose, nanocolumns are candidates for the ideal microstructure as it has been shown that when the diameters of these nanocolumns are optimized, resonant behavior is observed (10). Furthermore, nanocolumns offer a reduction in strain and defect states, and can improve flexibility and wear characteristics on the macro scale (11,12).  Similar benefits are observed in solid-state lighting technologies such as LEDs, with the addition of a dramatic increase in photoemission over that of planar films (13-16). Improvements in microstructure over previous work include the absence of secondary columns and a simplified growth process over etching techniques (16,17). These lead to improvements in optoelectronic properties -- a crucial property for photovoltaic devices.

Achieving uniform depositions of $In_xGa_{1-x}N$ needed for optoelectronic devices has proven challenging and can be attributed primarily to the difference in lattice spacing between InN and GaN (18-20). This lattice mismatch has resulted in phase segregation and decreases in crystalline and optical quality (21-23). Sophisticated deposition techniques such as molecular beam epitaxy (MBE), metal organic chemical vapour deposition (MOCVD), and hydride vapour phase epitaxy (HVPE) systems have achieved promising InGaN film depositions (9,17,24,25). Unfortunately these techniques are not economically scalable for large area (>1m$^2$) monolithic integration of PV. In addition, many of the best InGaN films were grown on expensive sapphire substrates. To overcome these challenges, this paper reports on studies using scalable plasma-enhanced evaporation deposition of $In_xGa_{1-x}N$ on an amorphous $SiO_2$ substrate where the ratio of metal source rates (indium/gallium) and substrate temperature were manipulated to investigate the effects on microstructure and composition to create nanocolumns appropriate for PV device integration.

**Experimental**

$In_xGa_{1-x}N$ thin films are deposited by a plasma-enhanced evaporation technique at a vacuum pressure of $10^{-4}$ Pa (prior to plasma introduction) and monitored in-situ by 30 kV reflective high energy electron diffraction (RHEED) (26,27). Silicon wafer (100) substrates have a 100 nm silicon dioxide coating grown with a tetraethyl orthosilicate precursor and are ultrasonically cleaned before being resistively heated at 600°C in the deposition chamber. A nitrogen plasma source (100 V, 45 A) is subsequently opened for 5 minutes while the substrate temperature ($T_s$), measured by infrared pyrometer, is lowered to 450°C to grow a 10 nm buffer layer of GaN to achieve adhesion. After RHEED shows the first signs of crystallinity in the GaN film via diffraction patterns, $T_s$ is increased to the deposition temperature, and the indium and gallium shutters are opened simultaneously. This yields the prescribed deposition rate, which is calibrated using a quartz oscillator. Deposition of the InGaN film is continued for 30 minutes to achieve nominal film thicknesses of 200 nm. Growth of the nanocolumns is believed to follow the self-catalyzing 'Ga balling' vapour-liquid-solid (VLS) mechanism reported in the literature for GaN and InGaN nanocolumns (7, 28-30). In this mechanism, Ga liquid droplets form on the $SiO_2$ substrate and establish liquid-solid interfaces. InGaN then selectively grows upwards from these GaN (Ga nitrided to GaN) islands forming nanocolumns.

By changing the temperature of the sources, the ratio of metal source rates (indium/gallium) were varied. This ratio and $T_s$ were manipulated to investigate the effects on microstructure and composition. Two $In_xGa_{1-x}N$ sample sets are examined in this work, in addition to a control sample: $T_s = 480$°C (A), $T_s = 540$ °C (B), and $T_s = 500$ °C (C). Within each sample set, the nominal indium source rate was altered from $x = 0.2$ to $x = 0.5$ in increments of $x = 0.1$, with sample number increasing with indium content. Sample C1 is pure GaN and is used as a control sample with an indium content of $x = 0$.

Electron microscopy imaging was conducted using a Leo Zeiss 1530 field emission scanning electron microscope (FESEM) and cross-section views were achieved by cleaving the samples with a diamond scribe.

The microstructure and lattice parameters were also examined using high resolution x-ray diffraction (XRD) focusing around the (0002) reflection. A Phillips X'Pert Diffractometer was used with a Cu-K$\alpha$ (1.54 Å) anode source operating at 50 kV and 30 mA. XRD data was fit using Gaussian profiles and the relative intensities were found by normalizing the InGaN (0002) reflection peak count to the background count for each sample.

The software program ImageJ was used to determine the void fraction of the InGaN nanocolumn layers as determined from top down SEM images. Thresholding, closing, filling and watershed processes were employed followed by a particle analysis.

**Results and Discussion**

Uniform films of $In_xGa_{1-x}N$ were successfully deposited as seen in Figures 1 and 2 and the microstructure was significantly influenced by the indium content of the samples. The FESEM images reveal distinct nanocolumns in the samples with lower indium incorporation (samples A1 and A2 and sample set B). The observed nanocolumns are hexagonal in geometry, which is expected due to the wurtzite structure of InGaN (17, 31). The columns are approximately normal to the substrate surface, show a high degree of order and display dimensional uniformity between individual columns. Since the evaporation technique is directionally tunable, another method of microstructural optimization is made available for future investigation. As the nominal indium content increases, the crystallinity decreases and a less ordered microstructure (losing distinct nanocolumns) is observed (samples A3 and A4). The widening of the nanocolumn diameter with indium content as seen in Figures 1 and 2 are consistent with the findings of Kuykendall et al. (7). In comparing materials with similar indium contents, an increase in average nanocolumn diameter was also noted for materials grown at the higher temperature (set B).

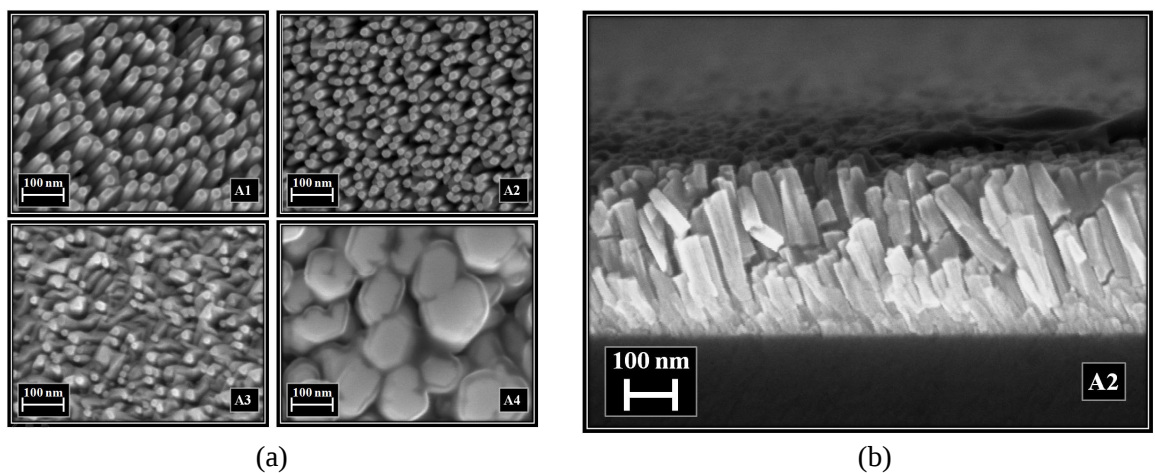

(a)          (b)

**Figure 1**: FESEM images showing **(a)** a decrease of crystalline order in microstructure with increasing indium content: A1 x = 0.21, A2 x = 0.32, A3 x = 0.36 and A4 x = 0.57, with all samples shown deposited at 480°C. The edge view in **(b)** shows a cross section of sample A2 as shown in **(a)** as cleaved with a diamond cutter taken perpendicular to the substrate surface. Note the broken columns are the result of cleaving and are not present in the undisturbed portions of the film. Average diameter of nanocolumns range from 15 nm to 25 nm.

The corresponding FESEM images for sample set B grown at the higher temperature of 540°C are shown in Figure 2:

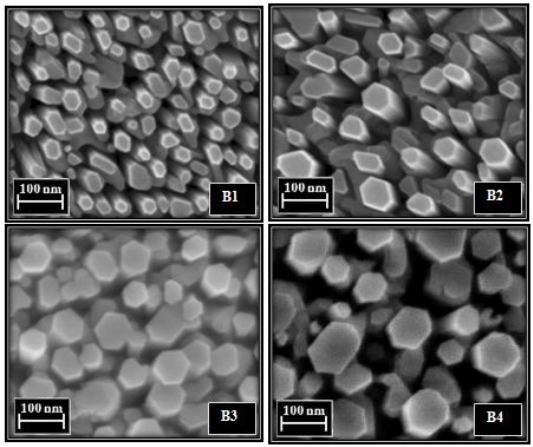

**Figure 2**: FESEM images showing increase in nanocolumn diameter with increasing indium content: B1 x = 0.08, B2 x = 0.24, B3 x = 0.29 and B4 x = 0.33, with all samples shown deposited at 540°C. Average diameter of nanocolumns range from 20 nm to 100 nm.

To verify the sample composition and provide a more quantitative view of the microstructure, XRD analysis was completed (Figure 3). Vegard's law was used to determine the indium content of each sample by taking a linear fit between the known c-lattice parameters of relaxed GaN (c = 5.185 Å) and InN (c = 5.703 Å) (32-34). In order to apply Vegard's law in this manner, the film must be relaxed and not under considerable strain (35). It is assumed that the films are sufficiently relaxed as the 200 nm film thickness is significantly larger than the critical layer thickness predicted by the model suggested by People and Bean as well as other experimental data in the literature for indium compositions ranging from x = 0 to x = 0.5 (36-38). Although it should be noted that this model was based on planar films.

As seen in Figure 3a, the XRD results show a clear shift in the (0002) InGaN peak, which indicates that the change in nominal source rates is being successfully reflected in the indium content of the samples. At the higher $T_s$ of 540°C less indium is being uniformly incorporated into the film than at 480°C. This is consistent with the literature and is a result of the higher temperatures increasing the desorption rate of indium atoms (24,39,40). The relative peak intensity of the XRD results seen in Figure 3b is an indication of uniformity of the nanocolumns' orientations. Sample set A shows higher XRD intensities than set B which is reflected by the increased density and order of sample set A (at similar indium levels). The FWHM, shown in Figure 3c, is an indication of crystallinity and the distribution of compositional phases of the sample. It also shows a decrease in crystallinity with samples grown at higher $T_s$, which is surprising as InGaN film crystal quality should improve with higher substrate temperatures given results elsewhere in the literature for InGaN films (18,24). This conflict can be explained by a difference in void fractions of the InGaN nanocolumn layer between the two sets of samples. The samples in set A (480 °C) show thinner but more closely packed nanocolumns compared to sample set B (540 °C). An ImageJ analysis of the SEM images determined that sample set A averages 23% fewer void spaces. Since there are fewer gaps for light to reflect off the amorphous sublayer during XRD measurements, the lower temperature samples (set A) would appear more crystalline even if the InGaN columns themselves are not. However, this reduction in void spaces may not fully explain the difference in peak widths and this anomalous trend warrants future investigation.

Another trend in the FWHM indicates that increased indium content reduces the crystallinity and increases the compositional modulation. This trend is also dependent on temperature, with the higher $T_s$ set displaying a steeper increase in FWHM with indium content. This trend has been previously observed by Ruterana and Deniel, who found that as the indium composition increased above 15%, a marked trend of decreasing crystallinity and increased phase segregation was observed (21). This trend can be attributed to the difference in atomic spacing between InN and GaN becoming a significant issue at higher indium/gallium ratios.

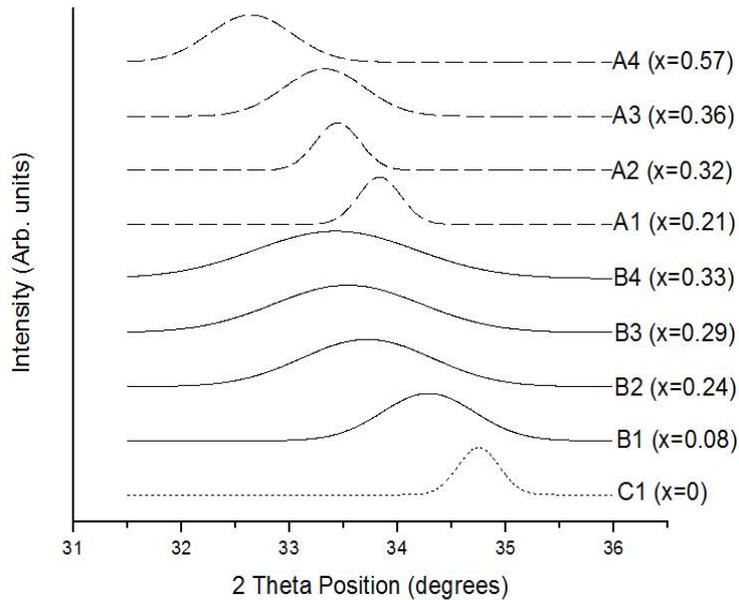

(a)

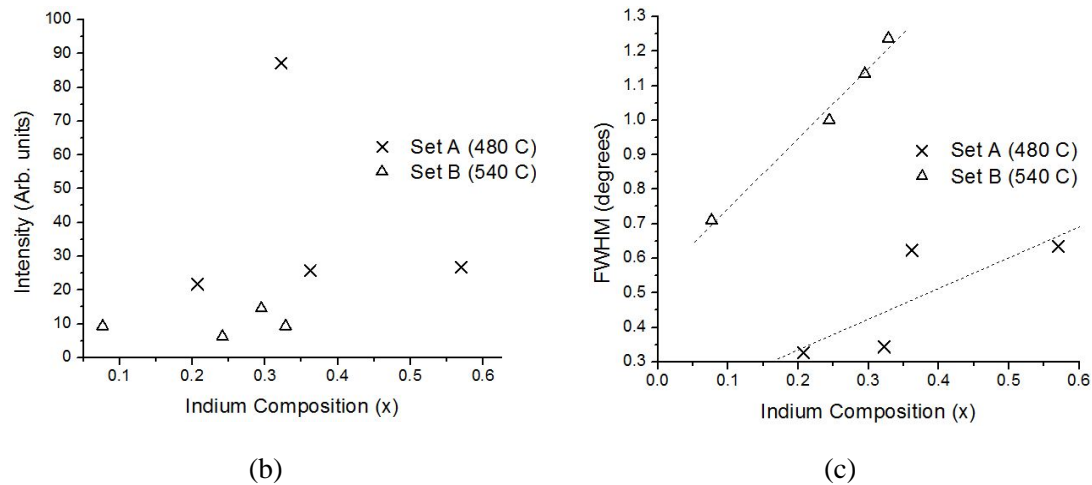

(b)                                              (c)

**Figure 3:** The results of XRD analysis show **(a)** the shift of the extracted (0002) reflection peak from a model fit to the collected data with indium composition, **(b)** the decrease in degree of crystallinity accompanying temperature as evidenced by the lower intensity of the (0002) reflection and **(c)** the broadening of compositional phase distribution with temperature as represented by the FWHM (dotted lines are only to guide the eye and to aid in distinguishing between sample sets).

No distinct phase segregation was observed, which could be a result of an adequate source of nitrogen radicals from the nitrogen plasma (41). Although it should be pointed out here, that measurements described here are not sensitive enough to draw conclusions about phase separation (depending on cluster sizes for instance, which could have composition fluctuations on the nm-scale). These results are promising for potential applications in optoelectronics, as the high degrees of crystallinity reduce non-radiative recombination centers and dislocations. The increased surface area provided by the nanocolumns offer opportunities for improved light output for solid-state lighting resulting in higher efficiencies (13-15). The oriented nanocolumnar structure also has benefits in simulations for photocarrier absorption and separation over planar solar photovoltaic structures (11).

**Conclusions**

This work has demonstrated the capability of a scalable deposition option for InGaN nanocolumn films for photovoltaic and solid-state lighting technologies. The control of the growth kinetics and composition of the films is critical for future developments that require specific microstructures and defined band gaps. These experiments have shown that nanocolumnar microstructures are favored at lower indium compositions and that lower temperatures have lead to thinner, but more closely packed nanocolumns. The high degree of crystallinity and unique geometry demonstrated in the nanocolumns grown offer interesting potential for solar photovoltaics and other optoelectronics devices.

**Acknowledgements**

The authors would like to acknowledge technical assistance from C. Elliott and T. Brown. This work was supported by the Natural Sciences and Engineering Research Council of Canada.

Published in: S. Keating, M.G. Urquhart, D.V.P. McLaughlin and J.M.Pearce, "Effects of Substrate Temperature on Indium Gallium Nitride Nanocolumn Crystal Growth", *Crystal Growth & Design*, **11** (2), pp 565–568, 2011.
http://dx.doi.org/10.1021/cg101450n**References**